# A Method for Discovering and Extracting Author Contributions Information from Scientific Biomedical Publications

Dominika Tkaczyk
ADAPT Centre
Trinity College Dublin
Ireland
Dominika.Tkaczyk@adaptcentre.ie

Andrew Collins
ADAPT Centre
Trinity College Dublin
Ireland
Andrew.Collins@adaptcentre.ie

Joeran Beel
ADAPT Centre
Trinity College Dublin
Ireland
Joeran.Beel@adaptcentre.ie

## ABSTRACT

Creating scientific publications is a complex process, typically composed of a number of different activities, such as designing the experiments, data preparation, programming software and writing and editing the manuscript. The information about the contributions of individual authors of a paper is important in the context of assessing authors' scientific achievements. Some publications in biomedical disciplines contain a description of authors' roles in the form of a short section written in natural language, typically entitled "Authors' contributions". In this paper, we present an analysis of roles commonly appearing in the content of these sections, and propose an algorithm for automatic extraction of authors' roles from natural language text in scientific publications. During the first part of the study, we used clustering techniques, as well as Open Information Extraction (OpenIE), to semi-automatically discover the most popular roles within a corpus of 2,000 contributions sections obtained from PubMed Central resources. The roles discovered by our approach include: experimenting (1,743 instances, 17% of the entire role set within the corpus), analysis (1,343, 16%), study design (1,132, 13%), interpretation (879, 10%), conceptualization (865, 10%), paper reading (823, 10%), paper writing (724, 8%), paper review (501, 6%), paper drafting (351, 4%), coordination (319, 4%), data collection (76, 1%), paper review (41, 0.5%) and literature review (41, 0.5%). Discovered roles were then used to automatically build a training set for the supervised role extractor, based on Naïve Bayes algorithm. According to the evaluation we performed, the proposed role extraction algorithm is able to extract the roles from the text with precision 0.71, recall 0.49 and F1 0.58.

## 1 INTRODUCTION

Authorship is an important concept in scholarly communication. It allows people to properly credit those who contributed to scientific discoveries and is widely used to assess people's scientific achievements. However, the fact that a person authored a paper gives only a partial picture. To fully evaluate the researcher's achievements, it is useful to know the nature of their contributions to the authored publications.

In some biomedical journals, authors are required to provide information about individual contributions, which is then attached to the manuscript in the form of a short section. These sections, typically entitled "Authors' contributions" (Figure 1) usually contain a natural language summary of individual authors' input to the described research and the publication itself. Examples of contributors' roles include: preparing the data, designing the experiments, writing the software or editing the manuscript.

**Authors' contributions**
AJ-M, HT, YS and RS carried out the IHC analysis. AJ-M, MN-M, MPU, V-MK and AM participated in study design and statistical analysis. AJ-M and MN-M drafted the manuscript. All authors read and approved the final manuscript.

**Figure 1: Example of "Authors' contributions" section in a biomedical publication with abbreviated author names.**

Unfortunately, these sections, as written in natural language, are unstructured and intended for humans rather than machines. Consequently, the analysis of this information requires time-consuming manual work, which makes processing large collections of documents expensive and impractical. We address these issues by proposing:

1. a method for discovering which roles are common in a corpus of sections of interest,
2. an algorithm to automatically extract the roles from unstructured text.

In the following sections, we describe the current state of the art, give details related to the methodology employed in our study, and finally report the study results.

## 2 RELATED WORK

Information extraction from scientific literature is a popular research area, resulting over the years in many approaches and tools. Tools for extracting information such as metadata or bibliographies from academic papers include CERMINE [1, 2, 3],

GROBID [4], PDFX [5], ParsCit [6], Science Parse[1] [7], Docear PDF Inspector [8] and OCR++ [9]. Comparisons of different approaches are also available [10, 11, 12]. However, none of the existing systems, to our best knowledge, is able to extract the information related to the contributions of individual authors directly from the content of the paper.

There is no globally accepted standard for scientific contributor roles set. Typically, authors just mention the roles they believe are important. There exists a small taxonomy of scientific contributor roles called CRediT[2]. CRediT was created manually and contains the following 14 roles:

- Conceptualization
- Data curation
- Formal analysis
- Funding acquisition
- Investigation
- Methodology
- Project administration
- Resources
- Software
- Supervision
- Validation
- Visualization
- Writing - original draft
- Writing - review & editing

Our study does not assume any input taxonomy but rather aims at discovering popular roles within a corpus of contribution descriptions in an unsupervised way. In Section 4.1 we briefly compare the results of our analysis to CRediT taxonomy.

Another study on PLOS authors contributions can be found on GitHub[3]. It contains an analysis of the distributions of roles and various statistics related to them. However, the authors do not analyze the descriptions written in natural language and operate only on a small predefined set of five roles.

## 3 METHODOLOGY

The study was based on a corpus of "Authors' contributions" sections obtained from PubMed Central Open Access Subset (PMC) resources[4]. The study was composed of two parts:

1. In the first part (Section 3.1), we used Open Information Extraction [13], as well as clustering techniques to analyze a corpus of contributions sections in order to discover which author roles are typical for biomedical publications.
2. In the second part (Section 3.2), we used the previously discovered author roles to train a machine learning algorithm able to extract roles from the text of a contributions section.

A typical contributions section contains several mentions of roles of individual authors. We represent these mentions as tuples containing:

- subject: "who", usually author name or initials,
- action: activity, often a verb phrase,
- object: "what the action was applied to", typically a noun phrase.

Figure 2 shows such decomposition of a single role mention. For example, the section in Figure 1 contains several role mentions. Represented as tuples, these role mentions are as follows: ("AJ-M", "carried out", "the IHC analysis"), ("MPU", "participated in", "statistical analysis"), ("All authors", "read", "the final manuscript"), ("All authors", "approved", "the final manuscript"), etc.

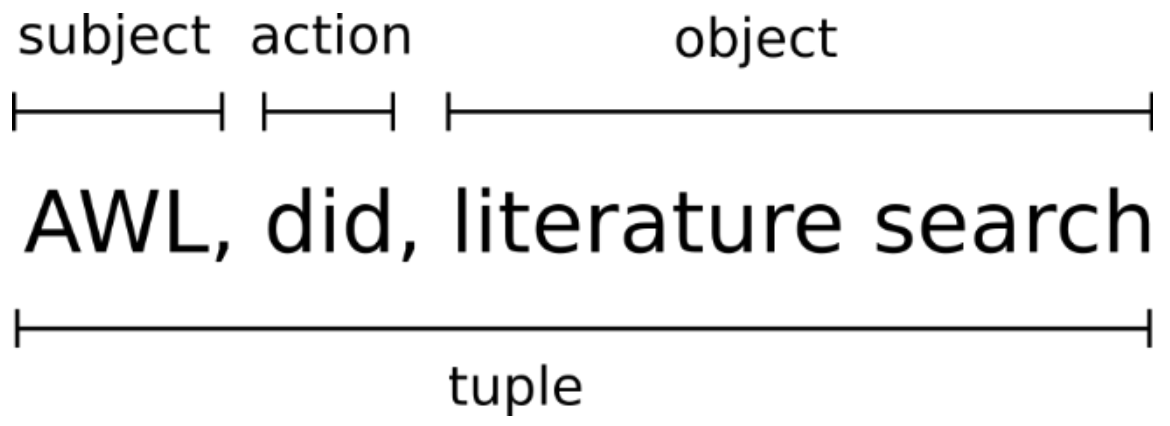

**Figure 2: The decomposition of a single role mention into three parts: subject, action, and object.**

It is important to note the difference between "role mentions" and "roles". Role mentions are parts of the raw text, while the roles can be understood as a finite set of abstract concepts. A role mention is an expression of a role in natural language. The same role can be expressed by many different role mentions. For example, the tuples: ("DT", "wrote", "the paper") and ("DT", "was responsible for", "writing the manuscript") express the same role ("paper writing") using different words (Figure 3).

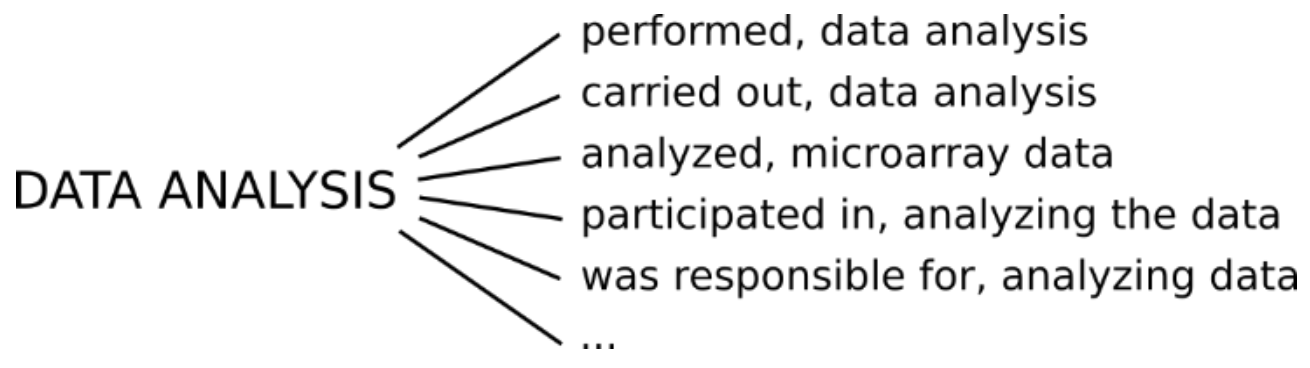

**Figure 3: An example of a role expressed in different ways by role mentions.**

### 3.1 Roles Discovery

In this section, we describe the analysis of a corpus of contributions sections, which was compiled using PMC resources. The main goal of this part of the study is to discover common roles appearing in the corpus. Our analysis is composed of the following stages:

1. Data preparation, where we gathered a corpus of contributions sections.
2. Data preprocessing:

[1] https://github.com/allenai/science-parse
[2] http://dictionary.casrai.org/Contributor_Roles
[3] https://github.com/hack4ac/plos-author-contributions
[4] https://www.ncbi.nlm.nih.gov/pmc/tools/openftlist/

a. Role mentions were extracted from the sections and redundant mentions were removed.
   b. Role mentions were then stemmed and stopwords were removed.
   c. Infrequent mentions were removed from the corpus.
   d. The role mentions were finally represented by a reduced number of important terms (keywords).
3. Finally, role mentions were clustered in order to discover roles (concepts).

Data Preparation

The data used in the analysis was obtained from PubMed Central Open Access Subset, which is a part of the total collection of articles in PMC, published under open licenses.

We downloaded the entire corpus of over 1.6 million documents in NLM JATS format[5]. NLM JATS is an XML-based format able to store machine-readable representation of a document, including rich metadata, formatting, hierarchy of sections along with their titles and paragraphs, table content, bibliography, etc. Within each downloaded document we searched for a section, whose normalized (lowercased and with all non-letters removed) title equals to “authorscontributions”. If such a section was found, we extracted its text, which contains the description of the authors’ roles. As a result, we obtained 186,874 descriptions of the authors’ roles written in English. For our study, we used a random subset of 2,000 such sections. The number of sections was reduced for performance reasons.

Preprocessing

At the beginning of our analysis, individual role mentions were extracted from the text. We used Stanford Open Information Extraction tool[6] for this task. OpenIE [13] is a comparatively new information extraction paradigm, in which it is possible to extract relations in the form of tuples (relation plus its two arguments) from the text, in an unsupervised way. The output corresponds roughly to role mentions described before, where action is the relation expression and subject and object are its two arguments.

As a result of applying OpenIE to our sections corpus, for every section we obtained a bag of role mentions, where a mention is a tuple of three text fragments from the original text. For example, from the sentence “AWL did the literature search and participated in the writing of the manuscript.” we got the following tuples: (“AWL”, “did”, “literature search”) and (“AWL”, “participated in”, “writing of manuscript”).

In practise, OpenIE tools tend to output tuples that are redundant. For example, from the same sentence we might get both (“authors”, “read”, “final manuscript”) and (“authors”, “read”, “manuscript”) tuples. We removed redundant tuples based on the comparison of actions and objects. More specifically, for each section we analyzed all pairs of tuples and removed one tuple from a pair if the following conditions were met: 1) their subjects are exactly the same, 2) the action of one tuple contains all the words of the other action in the same order, and 3) the object of one tuple contains all the words of the other object in the same order. This reduced the number of tuples without losing much information.

The roles in role mentions are expressed by action-object pairs, and the subject refers only to the author. At the beginning, our corpus of 2,000 sections contained 6,924 distinct action-object pairs, many of which expressed the same roles.

To merge some mentions and reduce the number of distinct action-object pairs, we applied cleaning and normalizing steps to actions and objects of role mentions. At the beginning, we stemmed the words within actions and objects and removed stopwords. For stemming we used R’s SnowballC library, and the stopwords list was downloaded from an online source[7]. This reduced the number of distinct roles to 6,289. After that, we removed infrequent role mentions, that is, mentions appearing less than five times in the corpus. This left us with only 434 distinct action-object pairs while keeping 55% of role mentions corpus.

**Table 1: Action and object keywords (common words) appearing in the corpus. The words are stemmed.**

| Action keywords | Object keywords |
|---|---|
| read, particip, draft, contribut, conceiv, perform, write, revis, carri, critic, approv, made, prepar, conduct, provid, review, supervis, equal, develop, edit, plan, initi, acquir, assist, coordin, help, took, undertook, gave, comment, take, recruit | manuscript, studi, data, final, design, analys, experi, collect, interpret, statist, respons, involv, paper, concept, result, version, substanti, acquisit, project, patient, research, work, content, intellectu, import, articl, discuss, first, protocol, molecular, investig, sequenc, literatur, idea, part, princip, clinic, trial, sampl, genet, laboratori, advic, tool |

Finally, we observed that due to splitting role mentions into two elements (action and object), we still have distinct mentions such as (“analys”, “data”) and (“perform”, “data analys”), which most likely refer to the same role. We wanted to normalize this, at the same time keeping the tuple-based structure of mentions. To achieve this, we extracted a number of most common terms from both actions and objects of the mentions (appearing at least 20 times in the corpus), and then each term was labeled as “action keyword” or “object keyword”, based on whether it is more common among actions or objects of the mentions (Table 1). Each role mention in the corpus was then transformed in the following way: 1) the subject was left intact, 2) all action keywords found in the original mention (in action or object) formed the new action, and 3) all object keywords found in the original mention (in action or object) formed the new object. In addition, if the new action turned out to be empty, we added a single "perform" keyword to it. This operation moved words between actions and objects so that action keywords are always in the actions of the mentions and object keywords are in their objects. For example, since "perform" is an action keyword, and "analys" and "data" are object keywords, the mention ("analys", "data") became ("", "data analys"), which brought it much closer to ("perform", "data analys"). This transformation left us with 285 distinct role mentions.

[5] https://jats.nlm.nih.gov/

[6] https://nlp.stanford.edu/software/openie.html

[7] http://www.ranks.nl/stopwords

It is important to note that our preprocessing is language specific. In order to adapt the analysis to other languages, we would need to provide alternative OpenIE tool, stemmer and stopwords list.

### Finding Roles

In this phase, we were interested in detecting roles in the role mentions corpus. Similarly as in standard ontology learning approach [14], we adopted unsupervised machine learning technique (clustering) for this task. Clusters of role mentions represent sought roles (concepts), in practice by enumerating common ways to express given role. Ideally, at the end of the clustering all mentions that refer to the same role, such as ("performed", "data analysis") and ("was involved in", "analyzing data"), should belong to the same cluster.

After preprocessing, our set contained 9,709 role mentions represented by cleaned subject-action-object tuples. We were interested in clustering the actions and the objects separately, which in result would define a third clustering based on the combinations of actions and objects.

More formally, let $M = \{m_1, \dots, m_N\}$ be the input mention set, and $A$ and $O$ be the set of action clusters and the set of object clusters (representing action and object concepts), respectively. We can define an action clustering as a function $f_a: M \to A$, which maps mentions to their action clusters. Similarly, let $f_o: M \to O$ be the mapping function which defines object-based clustering. This lets us define role set $R$ as a set containing all combinations of action and object concepts that share some mentions: $R = \{(a, o) \in A \times O \mid f_a^{-1}(a) \cap f_o^{-1}(o) \neq \emptyset\}$. Now the final combined clustering is a function $f_r: M \to R$ such that $\forall_{m \in M} f_r(m) = (f_a(m), f_o(m))$.

Set $R$ defines a binary relation between action and object clusters. We can define the weight of this relation as a number of the mentions the clusters share: $\forall_{a \in A, o \in O} r(a, o) = |\{m \in M \mid f_r(m) = (a, o)\}| = |f_r^{-1}(a, o)|$. Intuitively, if an action concept and an object concept appear in many role mentions together, they form a common role, and the weight of the role is large. This defines a graph structure among the clusters, with action and object concepts as nodes and weighted edges representing relation strength.

Finally, during our analysis we used the idea of a cluster label, defined as a bag of terms of the most numerous member of the cluster.

In general, we use a bottom-up clustering, where we start with initial action and object clusters and in several phases, we merge clusters together, effectively reducing their number. Initially, the clusters are defined as distinct normalized actions and objects. In other words, two mentions are in the same action/object cluster if and only if their normalized actions/objects are identical. Each round of clustering is composed of two stages. The first one is based purely on cluster term labels. The second one uses the graph structure defined previously. The clustering code was written in R. Algorithm 1 presents the pseudocode of the role mentions clustering.

**ALGORITHM 1:** Role mentions clustering

```
action_clusters ← grouping of actions by their normalized value
object_clusters ← grouping of objects by their normalized value
prev_similarity ← ∞
while prev_similarity < threshold do
  for each role_cluster_pair do
    if one element contains all terms of the other then
      merge clusters
      relabel the smaller cluster
    end
  end
  pair ← action or object cluster pair with the highest similarity
  similarity ← the highest similarity
  merge clusters from pair
  relabel the smaller cluster
end
```

The first stage of the clustering is based on the action/object label terms of the current role clusters. We examine pairs of role clusters and merge them if action and object terms of one of them contain the other cluster's terms. The new cluster is always given a label equal to the label of the bigger cluster from the examined pair.

The main clustering stage is based on the weighted graph relations between action and object clusters. First, we identify an action/object cluster pair that is most similar to each other, then their clusters are merged. When the highest similarity is below a predefined threshold, the clustering procedure terminates. We will only explain how the similarity between two action clusters is defined. The similarity between object clusters is defined analogously.

The main observation used for calculating the similarity between two action clusters is that two actions related to a lot of common objects will be more similar to each other. However, this assumption is trivially violated in cases where there simply are different ways we can affect the same object (for example the manuscript can be read, written, reviewed, etc.). In such cases we would like the overall similarity to be lower.

To reflect these observations, we introduce an object weight which is the reciprocal of the number of distinct actions it is related to: $\forall_{o \in O} w(o) = |\{a \in A \mid (a, o) \in R\}|^{-1}$. Intuitively, an object with a small weight (such as "manuscript") interacts with many different actions, in other words there are many actions that can be applied to it.

We define the similarity between two actions as the sum of the weights of all the objects they share: $\forall_{a_1, a_2 \in A} s(a_1, a_2) = \sum_{o \in O, (a_1, o) \in R, (a_2, o) \in R} w(o)$. Intuitively, two actions will have high similarity if: 1) they share a lot of objects, and 2) the objects they share are "specific" (not a lot of actions are applicable to them). In other words, a "broad" object, such that interacts with many actions, will not contribute much to the action similarity.

The clustering procedure resulted in reducing the number of role clusters from 285 to 63. The following clusters were merged:

1. "particip" and "perform"
2. "contribut" and "perform"
3. "assist" and "perform"
4. "manuscript" and "paper"
5. "project" and "study"
6. "carri" and "perform"
7. "experi" and "study"
8. "perform" and "undertook"
9. "manuscript" and "articl"
10. "approv" and "read"
11. "made" and "perform"
12. "conduct" and "perform"
13. "perform" and "supervis"
14. "help" and "perform"
15. "perform" and "plan"

The procedure made a few errors, merging for example: "approv" and "read", "perform" and "supervis". The final graph is shown in Figure 4.

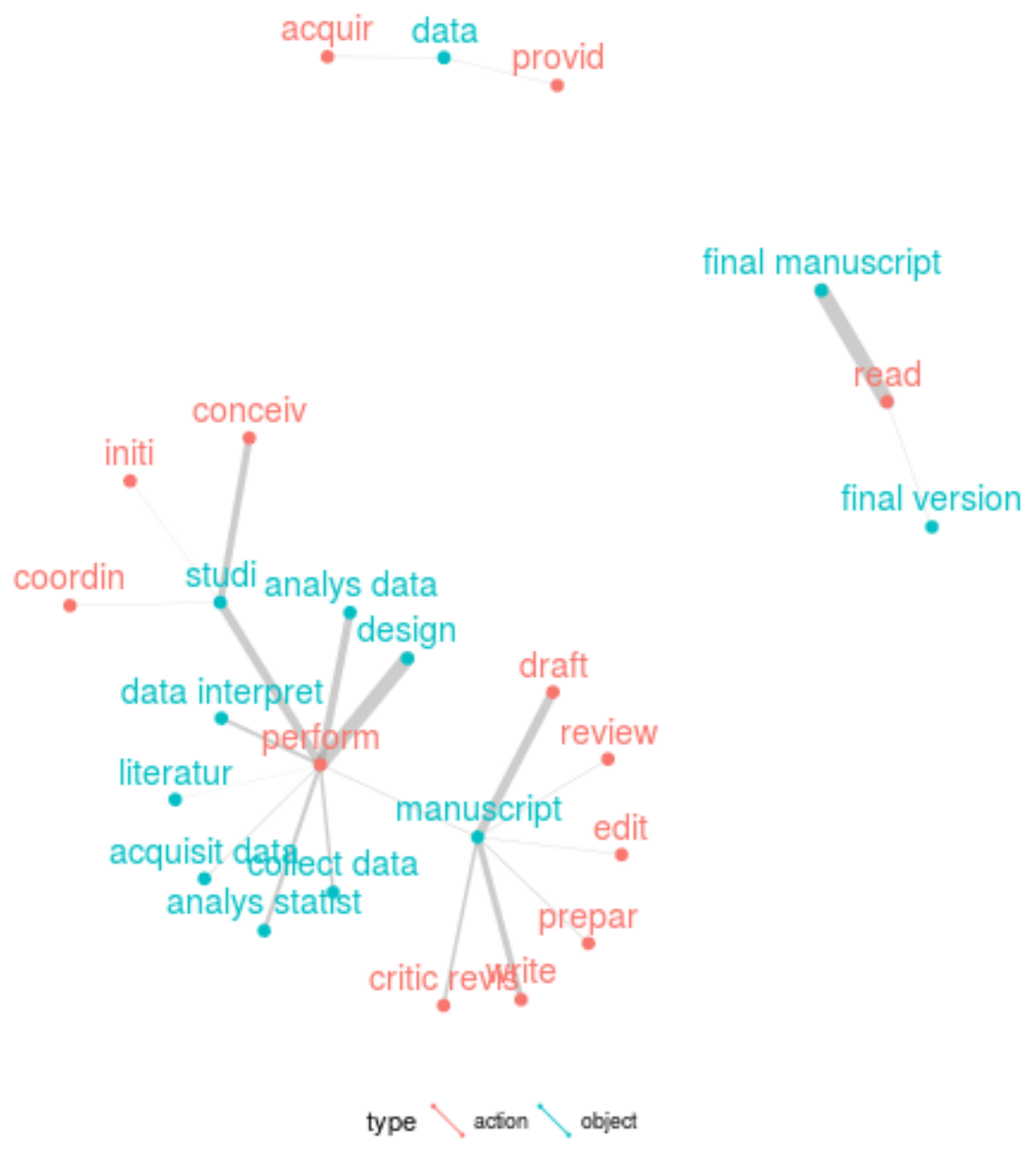

**Figure 4: The role graph resulting from automated clustering. The nodes represent action and object clusters (their labels are bags of stemmed terms). The width of edges represents the strength of the relation between action and object nodes. Less common roles were removed.**

The remaining role clusters were inspected manually. This included removing some clusters and merging others. Finally, we assigned manually the labels. The entire procedure resulted in the following 13 roles:

- Analysis
- Conceptualization
- Coordination
- Data collection
- Experimenting
- Interpretation
- Literature review
- Paper drafting
- Paper reading
- Paper review
- Paper revision
- Paper writing
- Study design

The final role set, as well as the fractions of mentions for every role, are presented in Figure 5.

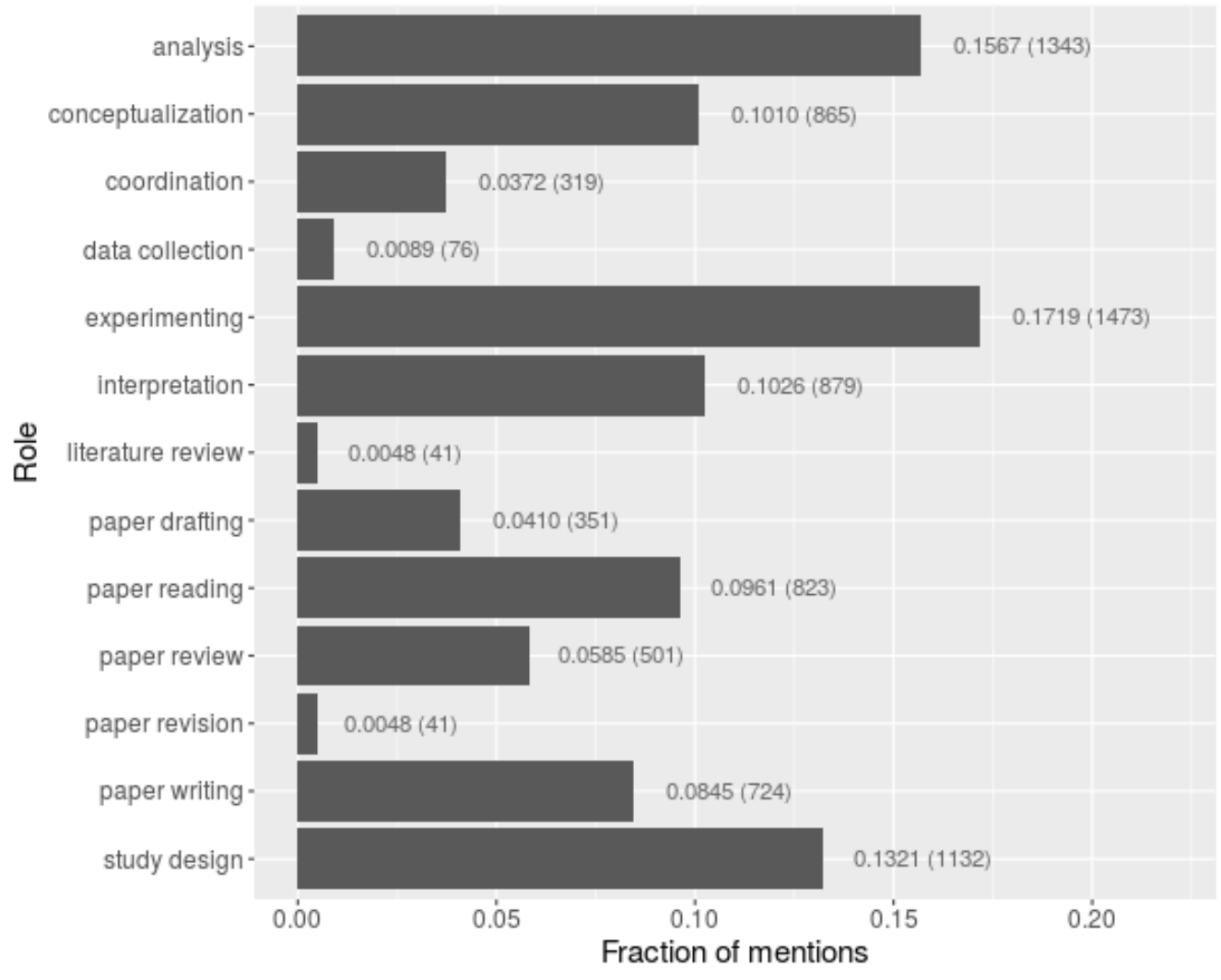

**Figure 5: The final set of roles, showing the counts and fractions of the entire role mention set.**

## 3.2 Roles Extraction

In this section, we describe our prototype of an automated extractor of authors' roles from the text. The extractor takes a description of authors' contributions on the input and outputs a set of extracted roles. We used previously developed preprocessing pipeline and discovered roles for this task.

The extraction algorithm is composed of the following steps:

1. First, OpenIE tools are used to extract a set of role mentions from the text.
2. Next, redundant mentions are removed.
3. From each remaining mention we extract 64 binary keyword-based features.
4. Each representation is then classified by a supervised model, resulting in one of the 13 discovered roles or a NULL value. NULL is returned if all feature values equal 0. These mentions are effectively discarded.

5. Finally, the remaining mentions are output as a set of subject-role pairs.

The supervised classifier is based on the results of the previous analysis. 64 binary features of a mention correspond to the presence of the object and action keywords extracted from the corpus before. As the classification algorithm we used Naïve Bayes from R's klaR library[8]. The training set was built automatically from cleaned clusters.

# 4 RESULTS

In this section we present the results of the evaluation of our proposed methods.

## 4.1 Roles Discovery

**Table 2: Comparison of the roles discovered by our study and existing taxonomy CRediT.**

| Our study | CRediT |
|---|---|
| Analysis | Formal analysis |
| Conceptualization | Conceptualization |
| Experimenting | Investigation |
| Study design | Methodology |
| Coordination | Project administration |
| Data collection | Resources |
| Paper drafting<br>Paper writing | Writing - original draft |
| Paper review<br>Paper revision | Writing - review & editing |
| Paper reading | - |
| Literature review | - |
| Interpretation | - |
| - | Software |
| - | Validation |
| - | Visualization |
| - | Funding acquisition |
| | Supervision |

In order to evaluate the results of the first part of our study, we compared manually the resulting clusters with the existing taxonomy CRediT. It is important to note that our study was performed using biomedical data only, while CRediT is a general-purpose taxonomy. Aa a result, some differences between the sets of roles are to be expected.

In general, the results are not very different (Table 2). Five roles appear in both our clusters and CRediT. Our study resulted in four roles related to preparing the manuscript itself, while CRediT has only two such roles. The missing roles in our results are either due to very low number of relevant mentions in the analyzed corpus or incorrect merging performed during the clustering.

## 4.2 Roles Extraction

In order to evaluate role extractor, we manually inspected 100 sections and provided subject-role pairs for them. 10 documents were removed due to the fact that they were not written in natural language, but rather contained a list of contributions in the following format (or a variation of it): "author1: role1, role2; author2: role3; ...". In such cases OpenIE tools fail to extract role mentions, and as a result we are currently unable to process them as well.

In the test set three new roles were discovered: paper approving, supervision and funding acquisition. Since the classifier does not have any training data for them, in the current version of our extractor those new roles are never assigned.

During the evaluation, for every document we compared the extracted subject-role pairs to the ground truth pairs. A pair was marked as correctly extracted if identical to any pair in the ground truth. This resulted in precision, recall and F1 scores for individual roles.

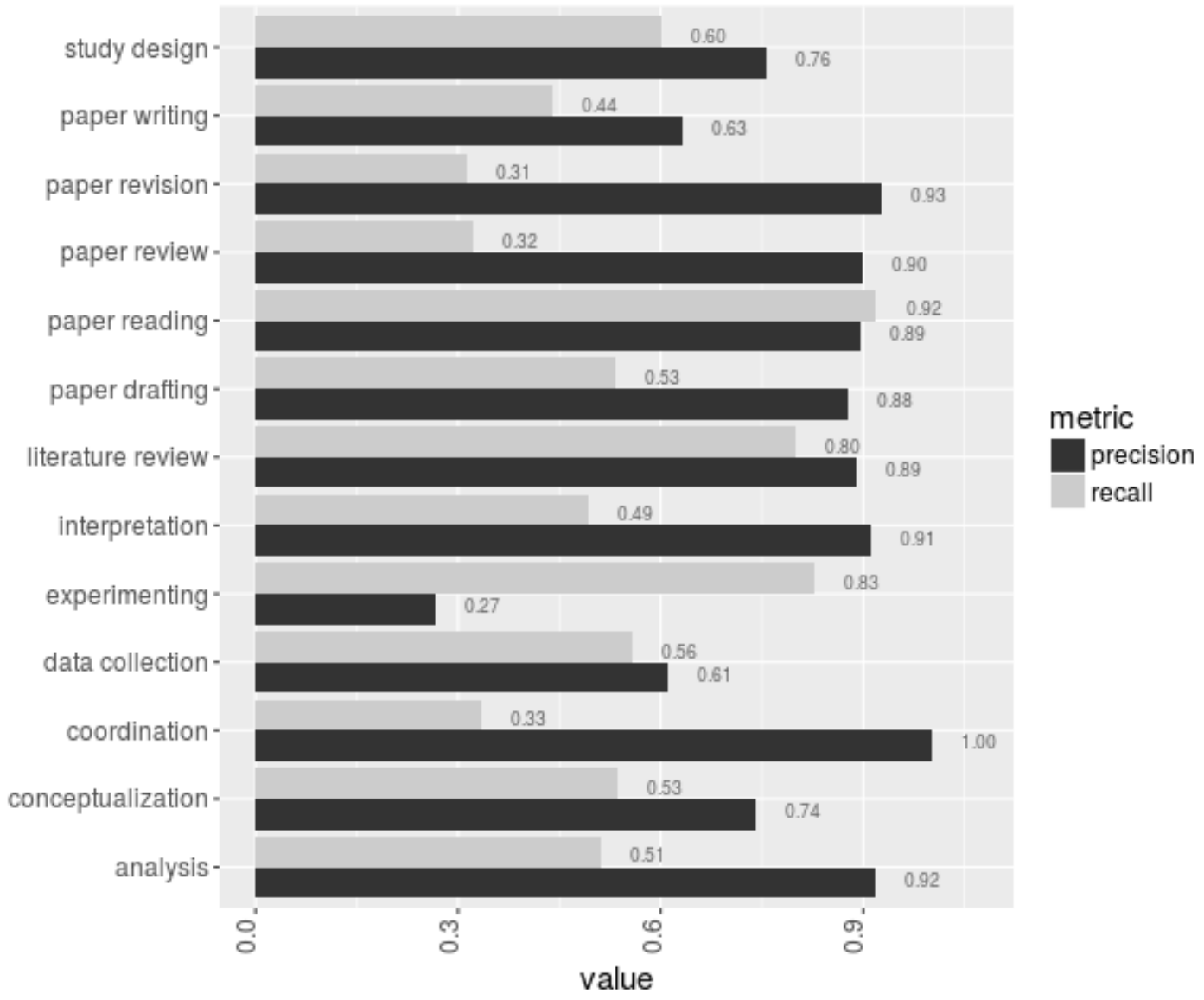

**Figure 6: Precision and recall for individual roles.**

[8] https://cran.r-project.org/web/packages/klaR

We obtained the following average results: precision 0.71, recall 0.49, F1 0.58. Figure 6 and Table 3 presents average precision, recall and F1 for individual roles.

**Table 3: Precision, recall and F1 for individual roles.**

| Role | Precision | Recall | F1 |
|---|---|---|---|
| Analysis | .92 | .51 | .66 |
| Conceptualization | .74 | .53 | .62 |
| Experimenting | .27 | .83 | .40 |
| Study design | .76 | .60 | .67 |
| Coordination | 1.0 | .33 | .50 |
| Data collection | .61 | .56 | .58 |
| Paper drafting | .88 | .53 | .66 |
| Paper writing | .63 | .44 | .52 |
| Paper review | .90 | .32 | .47 |
| Paper revision | .93 | .31 | .47 |
| Paper reading | .89 | .92 | .91 |
| Literature review | .89 | .80 | .84 |
| Interpretation | .91 | .49 | .64 |

## 4.3 Error Analysis

In order to have a clearer picture of the errors, we manually analyzed the mistakes made by the extractor in the test set. There are two types of mistakes:

- precision-related errors: a subject-role pair incorrectly present in the extracted output (results in lower precision)
- recall-related errors: a correct subject-role pair missing from the extracted output (results in lower recall)

We identified three main sources of errors (Figure 7):

- Errors related to mention extraction from the text. This happens when an incorrect mention is extracted or if a certain role mention is missing, and is responsible for 26% of precision errors and 73% of recall errors.
- Errors appearing during role discovery analysis, related to incorrect cluster merging. These errors result in the lack of three roles in the extractor's output and are responsible for 21% of recall errors.
- Classification errors, resulting in assigning an incorrect role to the tuple. These errors are responsible for 74% of precision errors and 6% of recall errors.

In general, the quality of the mention extraction stage has the biggest impact on the overall results, in particular recall. In a typical scenario, some mentions are missing from OpenIE output, which makes it impossible to extract specific subject-role pairs.

Incorrect tuples affect also the second error cause. For example, we observed that in many cases, Stanford's OpenIE tool extracts only one tuple from typical sentences similar to "All authors read and approve the final manuscript": ("all authors", "read", "the final manuscript"). Missing mention related to approving the manuscript resulted in the failure to discover this role in the corpus. In the future, we plan to experiment with different tuple extraction tools or approaches, which might give better results.

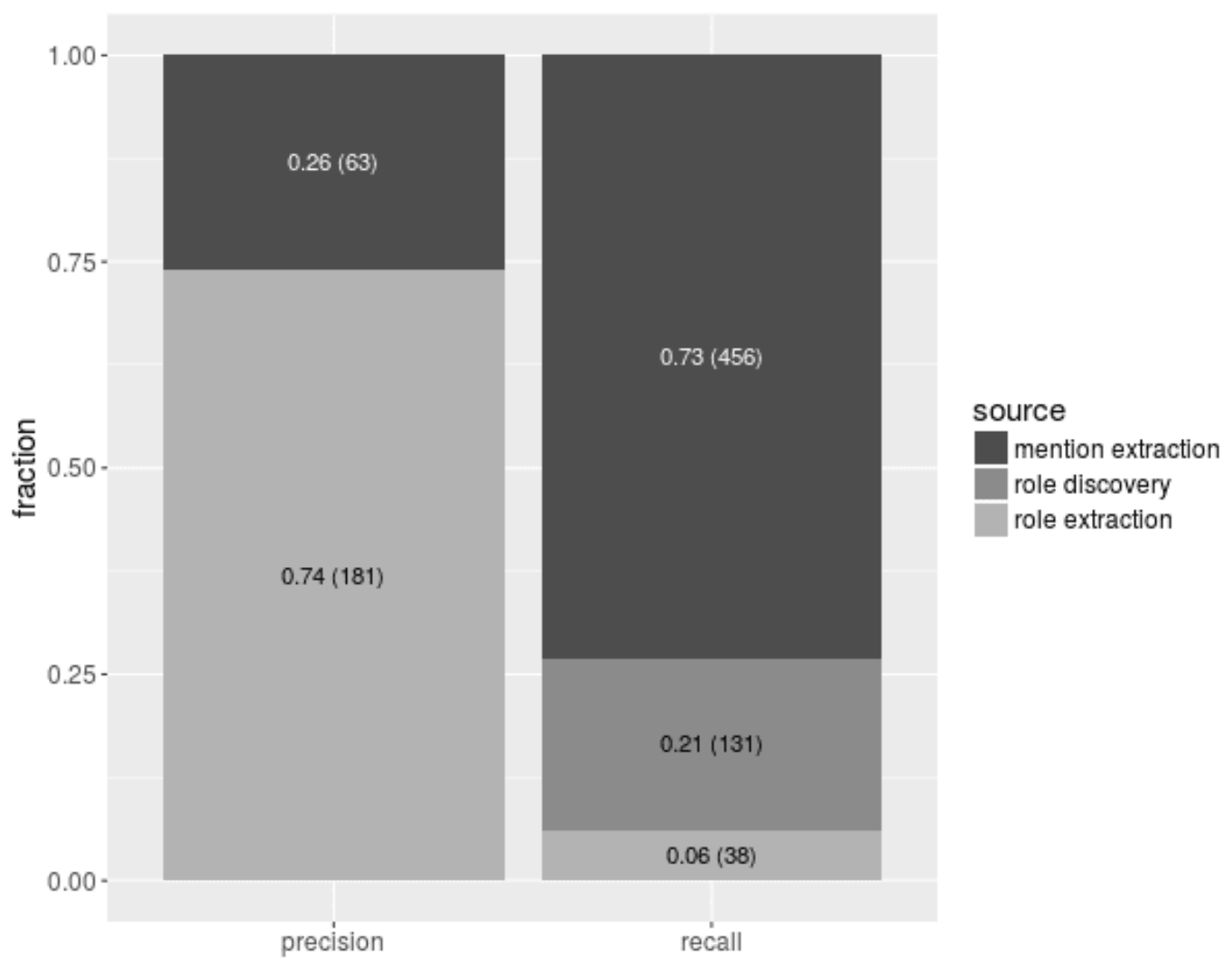

**Figure 7: The fraction of three error causes in types of errors (precision and recall errors).**

Finally, we observed that in many cases the classifier made the decision based on a single term such as "make", which obviously does not give enough information for a correct classification decision. This suggests that additional feature selection procedure for the classifier might result in better classification performance.

# 5 SUMMARY AND FUTURE PLANS

We presented a study of a collection of authors' contributions sections obtained from publications in biomedical disciplines. The results of our study include:

- a set of roles discovered in the data in an unsupervised manner,
- a prototype of a tool able to automatically extract the roles from the text of contributions sections.

The evaluation we performed shows the quality of both study parts. During the first part, we discovered the following roles: experimenting, analysis, study design, interpretation, conceptualization, paper reading, paper writing, paper review, paper drafting, coordination, data collection, paper review and literature review. The results of the unsupervised roles discovery are similar to existing taxonomy CRediT. Proposed automated role extractor is able to extract roles directly from the text with precision of 0.71, recall of 0.49 and F1 of 0.58.

Our study is a part of a larger effort related to releasing information and knowledge buried in millions of unstructured scientific documents and making it easily understood by the machines, which will contribute to solving the scientific information overload problem.

Our plans for the future include further improvement of the extraction methodology, including:

- trying out other tuple extraction approaches and tools,

- trying out a more interactive clustering approach, in which the user can prevent some clusters from merging,
- trying out alternative classification algorithms and feature selection approach,
- using coreference resolution to map subjects such as “he” or “they” to the original names of the authors.

We would also like to integrate the extraction tool into a larger system such as CERMINE. The new information about the individual authors’ roles could also be used to improve existing academic search engines [15, 16] and recommender systems [17, 18], including enriching the information displayed for the users and in researchers’ profiles.

## ACKNOWLEDGMENTS

This publication has emanated from research conducted with the financial support of Science Foundation Ireland (SFI) under Grant Number 13/RC/2106. The project has also received funding from the European Union’s Horizon 2020 research and innovation programme under the Marie Skłodowska-Curie grant agreement No 713567.